\documentclass[aoas,MSNbibl,nameyear,dvips]{arximspdf}
\usepackage{multirow}
\usepackage{graphicx}

\doi{10.1214/13-AOAS704} 
\volume{8}
\issue{1}
\pubyear{2014}
\firstpage{553}
\lastpage{576}

\makeatletter

\newtheorem{theorem}{Theorem}[section]
\makeatother

\begin{document}
\begin{frontmatter}

\title{Testing for shielding of special nuclear weapon~materials}
\runtitle{Testing for shielding}

\begin{aug}
\author[a]{\fnms{Kung-Sik} \snm{Chan}\corref{}\thanksref{t1}\ead[label=e1]{kung-sik-chan@uiowa.edu}},
\author[a]{\fnms{Jinzheng} \snm{Li}\ead[label=e2]{li\_jinzheng@yahoo.com}},
\author[b]{\fnms{William} \snm{Eichinger}\ead[label=e3]{william-eichinger@uiowa.edu}}\break
\and
\author[c]{\fnms{Erwei} \snm{Bai}\ead[label=e4]{erwei-bai@uiowa.edu}}
\thankstext{t1}{Supported in part by Department of Energy,
DE-FG52-09NA29364, PDP08-028.}
\runauthor{Chan, Li, Eichinger and Bai}
\affiliation{University of Iowa}
\address[a]{K.-S. Chan\\
J. Li\\
Department of Statistics\\
\quad and Actuarial Science\\
University of Iowa\\
Iowa City, Iowa 52242\\
USA\\
\printead{e1}\\
\phantom{E-mail:\ }\printead*{e2}}
\address[b]{W. Eichinger\\
Department of Civil\\
\quad and Environmental Engineering\\
University of Iowa\\
Iowa City, Iowa 52242\\
USA\\
\printead{e3}}
\address[c]{E. Bai\\
Department of Electrical\\
\quad and Computer Engineering\\
University of Iowa\\
Iowa City, Iowa 52242\\
USA\\
\printead{e4}}
\end{aug}

\received{\smonth{8} \syear{2012}}
\revised{\smonth{11} \syear{2013}}

%
\begin{abstract}
Nuclear-weapon-material detection via gamma-ray sensing is routinely
applied, for example, in monitoring cross-border traffic. Natural or
deliberate shielding both attenuates and distorts the shape of the
gamma-ray spectra of specific radionuclides, thereby making such
routine applications challenging. We develop
a Lagrange multiplier (LM) test for shielding. A strong advantage of
the LM test is that it only requires fitting a much simpler model that
assumes no shielding. We show that, under the null hypothesis and some
mild regularity conditions and as the detection time increases,  LM test statistic for (composite) shielding is
asymptotically Chi-square with the degree of freedom equal to the
presumed number of shielding materials. We also derive the local power
of the LM test. Extensive simulation studies suggest that the test is
robust to the number and nature of the intervening materials, which
owes to the fact that common intervening materials have broadly similar
attenuation functions. 
\end{abstract}

%

\end{frontmatter}

\section{Introduction}

A major homeland security issue concerns monitoring cross-border
traffic to prevent terrorists from smuggling nuclear weapon materials
into the US [\citet{Medalia}]. This requires screening cargoes which may
contain illegal radioactive materials, for example, special nuclear
weapon materials (SNWM).
A radioactive material generally gives off alpha, beta and gamma
emissions, and some material also gives off neutrons. Gamma ray spectra
are often used to detect specific nuclear materials, due to their
unique signature and the fact that gamma rays can travel relatively
long distances through most objects. Active detection methods [\citet
{Fetter1990,Moss2002}]
can detect quite small amounts of radioactive material, but they are
generally not preferable because they can be destructive and harmful to
humans. Consequently, passive sensing [\citet{August2005}] is the prime
detection approach.

In many applications (e.g., cargo screening), gamma-ray signature
recognition is plagued by the problem of weak signals, owing to (i)
relatively large distances between the radioactive source and the
detector (spectrometer) and (ii) natural or deliberate shielding by
intervening materials.
Furthermore, primary screening uses very low resolution detectors, and
higher resolution NaI detectors are used almost exclusively in
secondary vehicle screening when an alert is generated in primary screening.
Shielding not only attenuates the signal but it also distorts the
overall shape of the gamma ray spectrum, as the strength of attenuation
is specific to the energy (level) of the gamma ray and increases
exponentially with the thickness of the intervening material. Since the
nature and the thickness of the shielding materials are generally
unknown, shielding makes the problem of nuclide detection via gamma-ray
signature recognition challenging.

In the case of strong unshielded sources or long averaging times,
radionuclide detection can be carried out by peak finding methods
[\citet{Jarman}] and the library least-squares approach (LLS) [\citet
{Marshall1989,Gardner2009}]. The latter approach is based on the
principle that the pulse height (gamma ray) spectrum recorded by a
detector (the photon counts as a function of photon energy) is, on
average, equal to the weighted sum of the detector response functions
(DRF) to the gamma rays reaching the detector, plus that of the
background radiation, where the weights are the intensities of the
gamma rays. Consequently, the problem of deducing which radionuclides
are in the source may, in principle, be reduced to a regression
problem. Assuming no shielding, the relative intensities of the gamma
rays emitted by a radionuclide of interest are known so that their
detector response functions can be combined into one single detector
response function unique to the radionuclide, which makes the
regression problem more manageable.

In the general case of shielding, all gamma rays emitted by
radionuclides of interest and all the ways in which they may be
attenuated or modified must be included in the regression model.
Their DRFs may be modeled semiparametrically [\citet{Mitchell1989}] or
by Monte Carlo techniques as used in the Monte Carlo N-Particle code
(MCNP) [\citet{cashwell1957practical}] and GEometry ANd Tracking (GEANT)
[\citet{allison2006geant4}]. The library of all relevant DRFs is then
quite large, introducing multicollinearity and increasing the number of
parameters, which renders this approach not practical. This problem can
be solved by (i) grouping the detector response functions nuclide by
nuclide, (ii)~noting that the coefficients are all positive for a group
if the corresponding nuclide is present, and (iii) recognizing that
only a few radionuclides may be present in the source so that the
regression model is sparse, that is, most of the regression
coefficients are zero; see \citet{chan2011}, \citet{bai1} and \citet{kump}.

In principle, gamma ray recognition may be enhanced by incorporating
the physical laws of shielding. However, given the unknown nature of
the shielding material(s) of unknown thickness, a model-based approach
would require estimating a rather complex and nonlinearly parameterized
model. Here, we develop a Lagrange multiplier (LM) test for shielding
by known intervening materials which only requires fitting the much
simpler model under the null hypothesis of no shielding. Our approach
assumes that the mean background radiation (due to naturally occurring
radioactive sources, e.g., cosmic rays, granite, etc.) is known and we
are interested in detecting the presence of specific radionuclides, a
relatively easy task, in the absence of shielding. Thus, the critical
problem is to test for the presence of shielding, under the framework
of screening for specific radionuclides embedded in background
radiation. Solving this problem
is pertinent to achieving the real goal of detecting the known sources,
as detecting shielding would trigger human inspection.
We show that the LM test statistic has an asymptotic Chi-square
distribution under the null hypothesis of no shielding, with the degree
of freedom equal to the presumed number of shielding materials. The
attenuation patterns of commonly used intervening materials (e.g.,
carbon, lead, concrete and water) are broadly similar. It turns out
that this broad similarity results in the pleasant surprising
consequence that the power of the LM test appears to be robust to the
nature and number of common intervening materials, as suggested by our
simulation results. 

The paper is organized as follows. In Section~\ref{model} we introduce
the model for the pulse height spectrum under shielding. In Section~\ref{test} we derive the Lagrange multiplier test and its asymptotic null
distribution. The asymptotic power of the test is obtained in
Section~\ref{power}. Extensive simulations with shielding by simple or
composite intervening materials are reported in Section~\ref{simulation}.
Our approach focuses on attenuation of the gamma rays by intervening
materials but ignores the secondary effects of down scattering of the
gamma-ray spectra due to Compton scattering of the gamma rays outside
the detector, which introduces errors in the DRFs by changing the
relative contributions of each of the emissions from the radionuclide
and by changing the shape and intensity of the background radiation.
Indeed, background radiation generally varies over time, for example,
it is subject to vehicle suppression in cargo screening [\citet{lo2006}].
See \citet{Burr2009radio} for a review on the difficulties and
challenges of isotope identification using fielded NaI detectors.
Furthermore, DRF is detector-specific and may be unstable over time. In
Section~\ref{sec:sensitivity} we perform a sensitivity analysis of the
proposed method against errors in the DRFs and the background radiation.
We conclude in Section~\ref{conclusion} by pointing out some future
research necessary before field
deployment of the proposed methods.

It is desirable to test the proposed method with real data. However, to
access real data from any US port of entry, red tape has to be overcome and
the process has proved to be much harder than expected due to potential
national security issues.
The data used in the paper are semi-real in the sense that the
background was based on a real background and
isotope spectra were synthetically superimposed on the background
according to the standard and data published by the National Institute
of Standards and Technology (NIST).

\section{A Poisson regression model for shielding}\label{model}
Before presenting the statistical model for shielding, we outline the
physics underlying shielding of radionuclides; see \citet{Bryan2008},
\citet{Maher2008}, \citet{Hastings} and \citet{Cantone2011} for further
details. Each radionuclide emits a unique finite discrete set of gamma
rays, that is, the probability distribution of the emitted photon
energy is of a discrete type and unique to the radionuclide. For
example, Iodine-131 ($^{131}$I) emits five monoenergetic gamma rays at
energies 80.185 kiloelectron volt (keV), 284.31~keV, 364.489~keV,
636.99~keV and 722.911~keV, excluding those with branching ratio
(probability of emission) less than 0.5\%. The presence of a pure
intervening material reduces the intensity of a monoenergetic gamma ray
exponentially with the mass thickness of the intervening material (the
product of the thickness and density of the material), with a
multiplicative factor known as the attenuation coefficient, which is
specific to the energy of the gamma ray and the intervening material.

\begin{figure}

\includegraphics{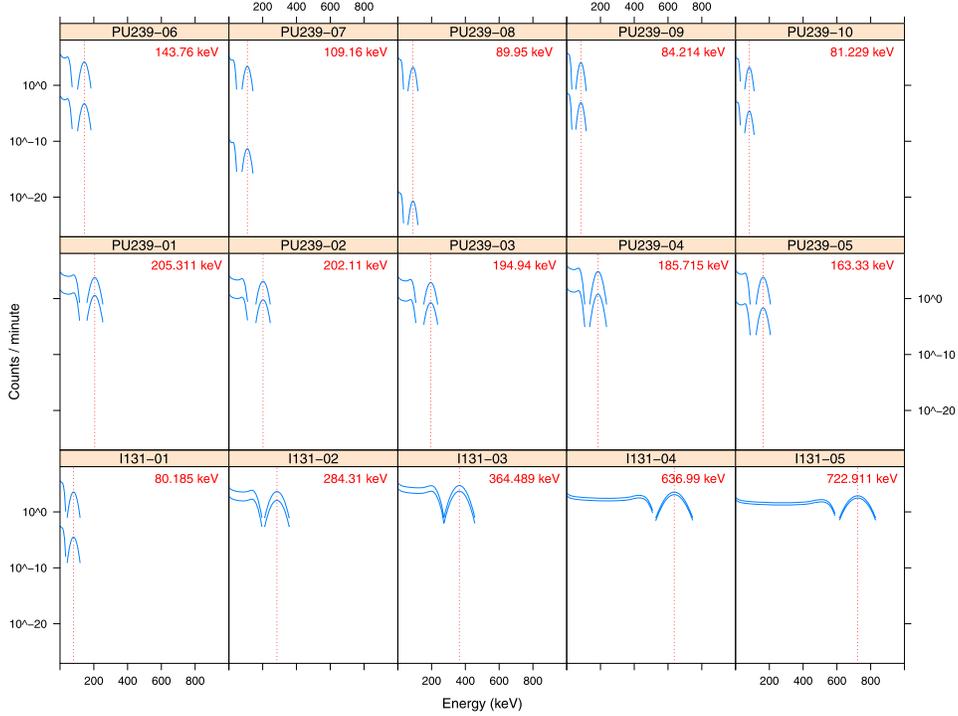}

\caption{Attenuated subspectra of monoenergetic gamma emissions for
$^{131}$I and $^{239}$Pu, on the logarithmic scale, whose branching
ratios (probabilities of emission) are greater than $0.5\%$, under lead
shielding with $x = 0$ (i.e., no shielding, upper curves) and $x=10$
(lower curves).}
\label{fig:leadSubspectra10Ch3}
\end{figure}

A monoenergetic gamma ray that reaches the detector leads to
scattering inside the detector, resulting in a distribution of photons.
The corresponding mean photon count as a function of the photon energy
is known as the detector response function, which is unique to the
energy level of the gamma ray entering into the detector. The spectrum
includes a response from energies other than that of the photon being
detected due to interactions of the photon inside the detector. The
upper curves in the bottom panel in Figure~\ref{fig:leadSubspectra10Ch3} display the detector response function (on
the logarithmic scale) of the five monoenergetic gamma rays emitted by~$^{131}$I.

We now present the statistical model for the pulse height spectrum
recorded in a detector. Let $Y_{i}$ represent the photon count at the
$i$th energy channel
of a spectrometer, where $i=1, \ldots, N$. [For example, $N=1024$
channels over the energy range up to 3 megaelectron volt (MeV).]
The photon counts are modeled as independent Poisson random variables,
with the mean of $Y_{i}$ given by
\[
\mu_{i}=\sum_{j=1}^{J}\sum
_{l=1}^{p_{j}}S_{ijl} a_{jl},
\]
where $ \{ S_{ijl}, i=1, \ldots, N \} $ is the detector response
function to the $l$th monoenergetic gamma ray
emitted by the $j$th nuclide, $l=1, \ldots, p_{j}, j=1, \ldots, J$.
For simplicity, background radiation is included in the library of
radionuclides postulated to be in the source, and it is indexed by $J$,
so $p_J=1$. Here, $a_{jl}$ denotes the
(relative) intensity of the $l$th monoenergetic gamma ray of the $j$th
nuclide.

Suppose that a slab of known intervening material of a pure type, for
example, lead, of thickness $t$
and density $\rho$ stands perpendicular to the path between a radioactive
source and the spectrometer.
The attenuation
of the gamma ray depends on its energy and the intervening material.
Let $I_{j0}$ be the initial emission intensity of the $j$th nuclei.
The intervening material then attenuates the intensity of its $l$th
monoenergetic gamma ray to $I_{jl}=I_{j0}\times e^{{-c_{jl}\times x}}$,
where $c_{jl}$ (cm$^2$/g) is the mass attenuation coefficient
(see the NIST websites\setcounter{footnote}{1}\footnote
{\url{http://physics.nist.gov/PhysRefData/XrayMassCoef/tab4.html}.}), and $x$
(g/cm$^2$) is the mass thickness, that is, the product of the
density $\rho$ and thickness $t$ of
the intervening material.
The attenuation formula can be extended to the case of multiple
intervening materials (e.g., carbon and lead), say, $M$ of them each of
mass thickness~$x_m$, $m=1,\ldots, M$:
\[
I_{jl}=I_{j0}\times e^{-\sum_{j=1}^M{c_{jlm}\times x_m}},
\]
where $c_{jlm}$ is the mass attenuation coefficient, due to the $m$th
intervening material, for the $l$th monoenergetic gamma ray
emitted by the $j$th nuclide. Here, we have omitted possible
down scattering of the gamma spectra by shielding materials of low
atomic number; see Section~\ref{conclusion}.


Consequently, in the presence of $M$ known intervening materials of
mass thickness $x_m, m=1,\ldots,M$, the photon counts over the $N$
energy bins, ${\mathbf{Y}}=(Y_1, \ldots, Y_N)^\top$, are modeled as
jointly independent Poisson random variables with the marginal
distribution given by
\[
Y_{i}\sim\operatorname{ Poisson } \Biggl(\mu_i = \sum
_{j=1}^{J} \sum_{l=1}^{p_j}S_{ijl}
 a_{jl} \Biggr),
\]
where
\[
a_{jl} = b_j \tau \exp \Biggl(-\sum
_{m=1}^{M}c_{jlm} x_m \Biggr),
\]
$b_j$ is the initial (unshielded) intensity of the $j$th nuclide per
unit time, and $\tau$ is the detection time which, for the moment, is
fixed. The background radiation (listed as the $J$th ``nuclide'' in the
library) is not shielded, hence, $c_{J1m}\equiv0$, for all $m=1,\ldots,M$.
Note the number of energy bins, $N$, is generally fixed.
However, the detection time can be increased without bound. So, the
asymptotic properties of the proposed method will be developed in the
framework of fixed $N$ but increasing~$\tau$.

\section{The Lagrange multiplier test}\label{test}
We consider the problem of testing for shielding of a
radioactive source that possibly comprises a finite number of known
radionuclides by $M$ known intervening materials of unknown mass
thickness.
The null hypothesis of no shielding is
\[
H_0\dvtx\ x_1 = x_2 = \cdots=
x_M = 0,
\]
that is, ${\mathbf{x} }=(x_1,\ldots,x_M)^\top= \mathbf{0} $, whereas the
alternative hypothesis of shielding is
\[
H_1\dvtx \mbox{ at least one of } x_m > 0,\qquad  m = 1, 2,
\ldots, M.
\]
A test for shielding
may proceed via a likelihood ratio test which requires fitting
the unshielded model under the null hypothesis and the general,
shielded model. The fitting of the latter model is quite complex, owing
to the nonlinear attenuation function. Thus, it is desirable
to develop a test that does not require estimating the
general model. Intuitively, under the
null hypothesis, the
constrained maximum likelihood estimator with $\mathbf{x}=0$
should be close to the
unconstrained maximum likelihood estimator, but otherwise they diverge
from each other. Hence, under the
null (alternative) hypothesis, the partial derivative of the
log-likelihood w.r.t.~$\mathbf{x}$, evaluated at
the constrained estimator, should be close to (diverge from) the zero vector.
The Lagrange multiplier test quantifies the
evidence against the null hypothesis in terms of some quadratic form
of the preceding partial derivative that
is designed to be
asymptotically $\chi^2$-distributed
under the null hypothesis.

We now elaborate the Lagrange multiplier
(score) test. First, we derive the score, the Hessian matrix and the
Fisher information.
Write\linebreak ${\bolds{\phi}}  = (x_1, x_2, \ldots, x_M, b_1,  b_2, \ldots,
b_J)^\top=({\mathbf{x} }^\top,\mathbf{b} ^\top)^\top$, where $ {\mathbf{x} } =
(x_1, \ldots, x_M)^\top$ and $\mathbf{b} = (b_1, b_2, \ldots, b_J)^\top$.
Let $u_{ijl}=S_{ijl}b_j \exp(-\sum_{m=1}^M c_{jlm} x_m)$,
$t_{ijl}=\break S_{ijl}\exp(-\sum_{m=1}^M c_{jlm} x_m)$ and
$U_i=\sum_{j=1}^J\sum_{l=1}^{p_j} u_{ijl}$. Note that the attenuation
coefficients $c_{jlm}$ are known for specified intervening materials.
The log likelihood function is given by
\[
L({\bolds{\phi} };\mathbf{Y}) = -\tau\sum_{i=1}^{N}U_i
+\sum_{i=1}^{N} Y_{i} \log(\tau
U_i) -\sum_{i=1}^{N}
\log(Y_{i}!).
\]

The score vector is given by
$\mathcal{U}_{\tau}({\bolds{\phi} }) = \nabla L({\bolds{\phi} };\mathbf{Y}) =
\bigl[
{ \mathcal{U}_{1,\tau} \atop \mathcal{U}_{2,\tau}}
\bigr]$, where
\[
\mathcal{U}_{1,\tau} = \biggl( \frac{\partial{L({\bolds{\phi} };\mathbf{
Y})}}{\partial{x_{m'}}}, m' = 1,
\ldots, M \biggr)^\top
\]
and
\[
\mathcal{U}_{2,\tau} = \biggl( \frac{\partial{L({\bolds{\phi} };\mathbf{
Y})}}{\partial{b_{k}}}, k = 1, \ldots, J
\biggr)^\top,
\]
with
\begin{eqnarray*}
\frac{\partial{L({\bolds{\phi} };\mathbf{Y})}}{\partial{x_{m'}}} &=& \tau\sum_{i=1}^{N}\sum
_{j=1}^{J} \sum_{l=1}^{p_j}
u_{ijl}c_{jlm'} \\
&&{}- \sum_{i=1}^{N}
Y_{i}U_i^{-1}{\sum
_{j=1}^{J} \sum_{l=1}^{p_j}
u_{ijl}c_{jlm'}},\qquad  m' = 1,\ldots, M
\end{eqnarray*}
and
\begin{eqnarray*}
\frac{\partial{L({\bolds{\phi} };\mathbf{Y})}}{\partial{b_{k}}}& =& - \tau\sum_{i=1}^{N}
\sum_{l=1}^{p_k} t_{ikl} \\
&&{}+ \sum
_{i=1}^{N} Y_{i}U_i^{-1}{
\sum_{l=1}^{p_k} t_{ikl} },\qquad k = 1,
\ldots, J.
\end{eqnarray*}
The second derivatives equal
\begin{eqnarray*}
\frac{\partial^2{L({\bolds{\phi} };\mathbf{Y})}}{\partial{x_{m''}}\,\partial
{x_{m'}}} &=& -\tau\sum_{i=1}^{N}
\sum_{j=1}^{J} \sum
_{l=1}^{p_j} u_{ijl} c_{jlm'}c_{jlm''}
\\
&&{}+\sum_{i=1}^{N} Y_{i} \Biggl
\{U_i^{-1} \sum_{j=1}^{J}
\sum_{l=1}^{p_j} u_{ijl}
c_{jlm'}c_{jlm''} \\
&&\hspace*{45pt}{}- U_i^{-2} \Biggl( \sum
_{j=1}^{J} \sum_{l=1}^{p_j}
u_{ijl} c_{jlm'} \Biggr) \sum_{j=1}^{J}
\sum_{l=1}^{p_j} u_{ijl}
c_{jlm''} \Biggr\}
\end{eqnarray*}
for $m', m'' = 1, \ldots, M$,
\begin{eqnarray*}
\frac{\partial^2{L({\bolds{\phi} };\mathbf{Y})}}{\partial{x_{m'}}\,\partial
{b_k}} &=& \tau\sum_{i=1}^{N}\sum
_{l=1}^{p_k} t_{ikl}c_{klm'}
\\
&&{}+\sum_{i=1}^{N} Y_{i} \Biggl
\{U_i^{-2} \Biggl( \sum_{j=1}^{J}
\sum_{l=1}^{p_j} u_{ijl}c_{jlm'}
\Biggr)\sum_{l=1}^{p_k} t_{ikl}
-U_i^{-1}\sum_{l=1}^{p_k}
t_{ikl} c_{klm'} \Biggr\}
\end{eqnarray*}
for $m' = 1, \ldots, M $ and $ k = 1, \ldots, J$, and
\[
\frac{\partial^2{L({\bolds{\phi} };\mathbf{Y})}}{\partial{b_h}\,\partial
{b_k}}= -\sum_{i=1}^{N}
Y_{i}U_i^{-2}{ \Biggl( \sum
_{l=1}^{p_k} t_{ikl} \Biggr) \Biggl( \sum
_{l=1}^{p_h} t_{ihl} \Biggr)}
\]
for $k, h = 1,\ldots,J$.

Let
\[
\mathcal{F}_{\tau} ({\bolds{\phi} })= -\nabla^2 L({\bolds{
\phi} };\mathbf{Y}) = \left[ %
\matrix{ \mathcal{F}_{11}
& \mathcal{F}_{12}
\vspace*{2pt}\cr
\mathcal {F}_{21} & \mathcal{F}_{22} }
\right]
\]
be partitioned according to $\mathbf{x}$ and $\mathbf{b}$,
where
\[
\mathcal{F}_{11} = \biggl(-\frac{\partial^2{L({\bolds{\phi} };\mathbf{
Y})}}{\partial{x_{m''}}\,\partial{x_{m'}}} \biggr),\qquad 
\mathcal{F}_{12} = \biggl(-\frac{\partial^2{L({\bolds{\phi} };\mathbf{
Y})}}{\partial{x_{m'}}\,\partial{b_k}}
\biggr),\qquad 
\mathcal{F}_{22} = \biggl(
-\frac{\partial^2{L({\bolds{\phi} };\mathbf{
Y})}}{\partial{b_h}\,\partial{b_k}} \biggr)
\]
and $\mathcal{F}_{21} = \mathcal{F}_{21} ^ \top$.

Consequently, the Fisher information matrix equals
\[
\mathcal{I}=\left[ %
\matrix{\mathcal{I}_{11} &
\mathcal{I}_{12}
\vspace*{2pt}\cr
\mathcal {I}_{21} & \mathcal{I}_{22} }
\right] = \tau\left[ \matrix{
 I_{11} &
I_{12}
\cr
I_{21} & I_{22}}
\right] = \tau I,
\]
where the arguments $m',m''$ below range from $1, \ldots, M$ and
$k, h = 1, \ldots, J$,
\begin{eqnarray*}
I_{11}&=& \biggl[ E \biggl(-\frac{1}{\tau}\frac{\partial^2{L({\bolds{\phi}
};\mathbf{Y})}}{\partial{x_{m''}}\,\partial{x_{m'}}}
\biggr) \biggr] \\
&=& \Biggl[ \sum_{i=1}^{N}
U_i^{-1} \Biggl( \sum_{j=1}^{J}
\sum_{l=1}^{p_j} u_{ijl}c_{jlm'}
\Biggr) \Biggl( \sum_{j=1}^{J} \sum
_{l=1}^{p_j} u_{ijl} c_{jlm''} \Biggr)
\Biggr],
\\
I_{21}^\top=I_{12}&=& \biggl[E \biggl(-
\frac{1}{\tau}\frac{\partial
^2{L({\bolds{\phi} };\mathbf{Y})}}{\partial{x_{m'}}\,\partial{b_k}} \biggr)
\biggr]\\
& =& \Biggl[-\sum
_{i=1}^{N} U_i^{-1} \Biggl(
\sum_{j=1}^{J} \sum
_{l=1}^{p_j} u_{ijl}c_{jlm'}\Biggr)
\Biggl( \sum_{l=1}^{p_k} t_{ikl}
\Biggr) \Biggr]
\end{eqnarray*}
and
\[
I_{22}= \biggl[E \biggl(-\frac{1}{\tau}\frac{\partial^2{L({\bolds{\phi}
};\mathbf{Y})}}{\partial{b_h}\,\partial{b_k}} \biggr)
\biggr]= \Biggl[ \sum_{i=1}^{N}
U_i^{-1} \Biggl( \sum_{l=1}^{p_k}
t_{ikl} \Biggr) \Biggl(\sum_{l=1}^{p_h}
t_{ihl} \Biggr) \Biggr].
\]

Let $\mathcal{U}_{1,\tau}(\tilde{{\bolds{\phi} }};\mathbf{Y}) $ be an
$M$-dimensional column vector with its $m$th element equal to $ \frac
{\partial{L(\tilde{{\bolds{\phi} }};\mathbf{Y})}}{\partial{x_{m}}}$.
Let $\tilde{{\bolds{\phi} }}=(\mathbf{0} ^\top,\tilde{\mathbf{b} }^\top)^\top
=(0,\ldots,0,\tilde{b}_1,\ldots\tilde{b}_J)^\top$ be the constrained
maximum likelihood estimator of ${\bolds{\phi} }$ under the null hypothesis.
Note that $\tilde{\mathbf{b} }$ can be computed via the EM algorithm. Let
$I^{11}$ be the $(1,1)$ block of the inverse of the Fisher information
matrix $I$. The Lagrange multiplier (LM) test statistic equals
\begin{eqnarray*}
\xi_{\mathrm{LM}} &=& \frac{1}{\tau}{\mathcal{U}_{1,\tau}(\tilde{{
\bolds{\phi} }};\mathbf{Y})}^\top I^{11}(\tilde{{\bolds{\phi}
}}) \mathcal{U}_{1,\tau
}(\tilde{{\bolds{\phi} }};\mathbf{Y})
\\
&=& \frac{1}{\tau}{\mathcal{U}_{1,\tau}(\tilde{{\bolds{\phi} }};
\mathbf{ Y})}^\top{ \bigl\{I_{11}(\tilde{{\bolds{\phi}
}})-I_{12}(\tilde{{\bolds{\phi } }})I^{-1}_{22}(
\tilde{{\bolds{\phi} }})I_{21}(\tilde{{\bolds{\phi} }}) \bigr
\}}^{-1} \mathcal{U}_{1,\tau}(\tilde{{\bolds{\phi} }};\mathbf{Y}).
\end{eqnarray*}

Below we show that the LM test statistic is asymptotically $\chi^2$
distributed and that the classical result on the asymptotic equivalence
between the LM test, the likelihood ratio test and the Wald test holds,
under general regularity conditions. Recall the likelihood ratio (LR)
test statistic equals
\[
\xi_{\mathrm{LR}} = 2 \bigl\{L(\hat{\bolds{\phi} })-L(\tilde{\bolds{\phi} })
\bigr\},
\]
where $\hat{\bolds{\phi} } =(\hat{\mathbf{x} }^\top,\hat{\mathbf{b} }^\top
)^\top$ denotes the unconstrained ML estimator under the general hypothesis,
and the Wald test statistic equals
\begin{eqnarray*}
\xi_{\mathrm{W}} &=& \sqrt{\tau}\hat{\mathbf{x} }^{\top} \bigl
\{I^{11}(\hat{\bolds{ \phi} }) \bigr\}^{-1}\sqrt{\tau}\hat{
\mathbf{x} }
\\
&=& \tau\hat{\mathbf{x} }^{\top} { \bigl\{ I_{11}(\hat{\bolds{
\phi} })-I_{12}(\hat{\bolds{\phi} }) I^{-1}_{22}(
\hat{\bolds{\phi} })I_{21}(\hat {\bolds{\phi} }) \bigr\}} \hat{
\mathbf{x}}.
\end{eqnarray*}

The following assumption is required for the asymptotic equivalence of
the preceding tests under the null hypothesis:
\begin{longlist}[(A1)]
\item[(A1)] The true mean unshielded spectrum is a positive function,
that is, $\sum_{j=1}^J b_j S_{ij\bullet}>0$ for every energy
channel $i=1,\ldots,N$, where $S_{ij\bullet}=\sum_{l=1}^{p_j} S_{ijl}$. Denote
the partial derivatives of the mean spectrum w.r.t. $\mathbf{x}$
and $\mathbf{b}$, evaluated at $\mathbf{x}=0$ and
at the true $\mathbf{b}$, by
\[
\mathbf{v} _q = \Biggl(-\sum_{j=1}^{J}
b_j\sum_{l=1}^{p_j}
S_{ijl} c_{jlq}, i =1,\ldots, N \Biggr)^\top
\]
for $q = 1,\ldots, M$ and
\[
\mathbf{v} _{q+M} = ( S_{iq\bullet}, i =1, \ldots, N
)^\top
\]
for $q = 1, \ldots, J$. Then the $M+J$ $\mathbf{v}$'s are linearly independent.
\end{longlist}

Note that $U_i$ is the mean of $Y_i$ at energy channel $i$. Under the null
hypothesis of no shielding, $U_i=\sum_{j=1}^J b_j S_{ij\bullet}$, so
({A1}) states that the mean unshielded gamma-ray spectrum is a
positive function. As a function of $i$, $S_{ij\bullet}$ is the overall
DRF for the $j$th nuclide in the source, while $-\sum_{j=1}^{J} b_j\sum_{l=1}^{p_j} S_{ijl} c_{jlq}$ is the
instantaneous reduction rate in the mean unshielded gamma-ray spectrum
per unit
mass thickness of the $q$th intervening material, evaluated under the
hypothesis of no shielding. The linear independence assumption in ({A1}) thus
formalizes the requirements that distinct radionuclides have
unique gamma-ray signatures and distinct intervening materials admit
unique attenuation patterns.

\begin{theorem}\label{Thm:chisq}
Assuming \textup{({A1})} and as the detection time $\tau\to\infty$, the LM
test statistic, the Wald test statistic and the likelihood ratio test
statistic all have asymptotic $\chi^2_M$ distributions under $H_0$:
$\mathbf{x} = \mathbf{0}$.
\end{theorem}
We defer the proofs of all theoretical results to the supplemental
article \citet{chan2013suppl}. Arguably, ({A1}) is a mild assumption even
though the second part fails if the attenuation functions (the
attenuation coefficient as a function of the energy of the gamma ray)
of the $M$ intervening materials are linearly dependent. Indeed, $M$
intervening materials with broadly similar attenuation functions may be
practically approximated by one of the intervening materials with a
modified mass thickness. Below, we illustrate that common intervening
materials have broadly similar attenuation functions. Extensive simulation
results reported below suggest that the LM test is robust to the nature
and the number of intervening materials.

\section{Local power of the LM test}\label{power}
In this section we derive the local power of the Lagrange multiplier
test, the likelihood ratio test and the Wald test for testing
\[
H_0 \dvtx \bolds{\phi} = \bolds{\phi} _0
\]
versus
\[
H_{1,\tau}\dvtx \bolds{\phi} = \bolds{\phi} _0 +
\frac{\mathbf{h} }{\sqrt{\tau}},
\]
where $\mathbf{h} = (\mathbf{h} _1^\top, \mathbf{h} _2^\top)^\top$, with the
vector partition being similar to that of $\bolds{\phi}$.
The following result shows that the tests reject a fixed alternative
with probability approaching 1 as $\tau\to\infty$.

\begin{theorem}\label{Thm:power}
Assume \textup{{(A1)}} holds. Then as $\tau\to\infty$, the Lagrange
multiplier test statistic, the likelihood ratio test statistic and the
Wald test statistic have noncentral $\chi^2_M$ distributions with the
noncentral parameter equal to $\mathbf{h} _1^\top
(I_{11}-I_{12}I_{22}^{-1}I_{21})\mathbf{h} _1$ under $H_{1,\tau}\dvtx \bolds{
\phi} = \bolds{\phi} _0 + \frac{\mathbf{h} }{\sqrt{\tau}}$, where $ \mathbf{h}
= (\mathbf{h} _1^\top, \mathbf{h} _2^\top)^\top$.
\end{theorem}
%

\section{Simulation studies}\label{simulation}
The modeled detector characteristics are those of a $3\times3$ inch
cylindrical NaI detector whose spectral response is degraded to a
resolution at the
Cesium 662 keV  peak of 80~keV full width half maximum (FWHM) of the peak;
consequently, the resolution is smaller at low energies, similar to
real measurements. The resolution of the detector is degraded from that
of a standard NaI detector to better approximate the response of a
larger detector or different scintillation material that may be more
appropriate for a drive through a portal system. The unshielded DRFs
are constructed by assuming each source to be small and pure, placed at
an arbitrary distance from the flat face of the detector with no
materials intervening between the source and detector. This ignores the
effects of attenuation and matrix effects in the source, but allows one
to concentrate on identification in situations with poor detector resolution.
The mean of the background signal used for the simulation is based upon
a long-duration, measured background using a NaI detector at the US
Military Academy Accelerator facility. This background is adjusted to
include radiation from terrestrial sources pertaining to the $^{238}$U
decay chain (e.g., $^{214}$Pb, $^{214}$Bi, $^{208}$Tl) that are
degraded to the same resolution as the modeled detector; see \citet
{mitchell2009skyshine}. The data are further adjusted so that the
variability of the counts about the long-term mean is consistent with a
short duration measurement ($\sim$1 minute). The simulation of the
possibly shielded data included this background and further assumed
that there are two radionuclides in the source, namely, $^{131}$I and
$^{239}$Pu, whose signatures are simulated as described above.
$^{239}$Pu is chosen because it is the primary fissile isotope used
for the production of nuclear weapons and the emissions from $^{239}$Pu
are quite weak and easily attenuated since many of the lines are of
relatively low energy. $^{131}$I is chosen because it is fairly common
(and thus may not trigger a more extensive search) and has low energy
emission lines that might be used to intentionally mask the signature
from $^{239}$Pu. This scenario represents a possible method to hide the
transport of $^{239}$Pu.
The response of the detector is calculated for each emission line for
each of the radionuclides separately. These detector response functions
will also be referred to as the subspectra for each of the nuclei. The
detector response includes such phenomena as photoelectric absorption,
pair production, Compton scattering, backscattered photons, escape
peaks and detector efficiency. Each of the subspectra is calculated in
proportion to the branching ratios of the emission lines. Thus, in the
absence of shielding, the mean total response for a given radionuclide
is obtained by multiplying all of the constituent subspectra by the
same constant and adding the resultant spectra.
Notice that all the subspectra of $^{239}$Pu are under 300~keV and so
are some subspectra of $^{131}$I.

\begin{figure}

\includegraphics{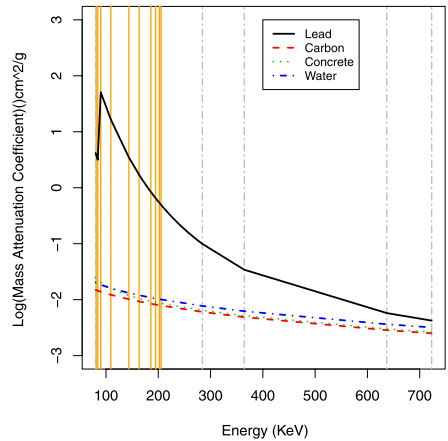}

\caption{$^{131}$I and $^{239}$Pu mass attenuation coefficients, with
vertical lines corresponding to the peak energy of each subspectrum of
$^{239}$Pu and vertical dot-dashed lines corresponding to the peak
energy of each subspectrum of $^{131}$I.}
\label{fig:I131Pu239mac}
\end{figure}

Four common intervening materials, namely, carbon, concrete, lead and
water are of interest. A plot of the mass attenuation coefficients (on
log scale) against the energy of the monoenergetic gamma ray, under
water, concrete, carbon and lead shielding for the two radionuclides
$^{131}$I and $^{239}$Pu are presented in Figure~\ref{fig:I131Pu239mac}. (The mass attenuation coefficients are obtained by
spline interpolation based on the tables of mass attenuation coefficients
posted on the NIST website.%
\footnote{\url{http://physics.nist.gov/PhysRefData/XrayMassCoef/},
with the path for
water shielding:
ComTab/water.html;
concrete shielding:
ComTab/concrete.html;
carbon shielding:
ElemTab/z06.html;
lead shielding:
ComTab/glass.html.}) The curves of mass attenuation coefficients for
carbon, water and concrete shielding are quite similar. Lead has higher
attenuation than the other three materials, at any energy level of the
photon, but it also takes a broadly similar attenuation pattern.

\subsection*{Simple intervening material}

We first consider the case of a single intervening material, in which
case simulation is done as follows:
\begin{longlist}[1.]
\item[1.] Let $b_{1}=1$ be the coefficient of $^{131}$I, $b_{2}=0.15$ the
coefficient of $^{239}$Pu, and $b_{3}=1$ that of the
background.
\item[2.] Determine $c{}_{1l}$ and $c{}_{2l}$ (cm$^2$/g), where $c{}_{1l}$
and $c{}_{2l}$ are the mass attenuation coefficients under the
specified intervening material for each monoenergetic gamma ray of
$^{131}$I and $^{239}$Pu in the library.
Note the background is not attenuated by the shielding material so
$c{}_{31}=0$.
\item[3.] Calculate $a_{1l} = b_{1}  \tau\exp(-c_{1l}\times x)$ and
$a_{2l} = b_{2}  \tau\exp(-c_{2l}\times x)$ for each
$x$, respectively, where $\tau= 1$.
\item[4.] Simulate $Y_{i}$ as independent Poisson random variables with
mean\break $\sum_{j=1}^{J}\sum_{l=1}^{p_{j}}S_{ijl} a_{jl}$.
Note $J=3$ since we only have $^{131}$I, $^{239}$Pu and background.
\end{longlist}
%
The parameter $x$ ranges from 0 to 30 (g/cm$^2$). 

\begin{figure}

\includegraphics{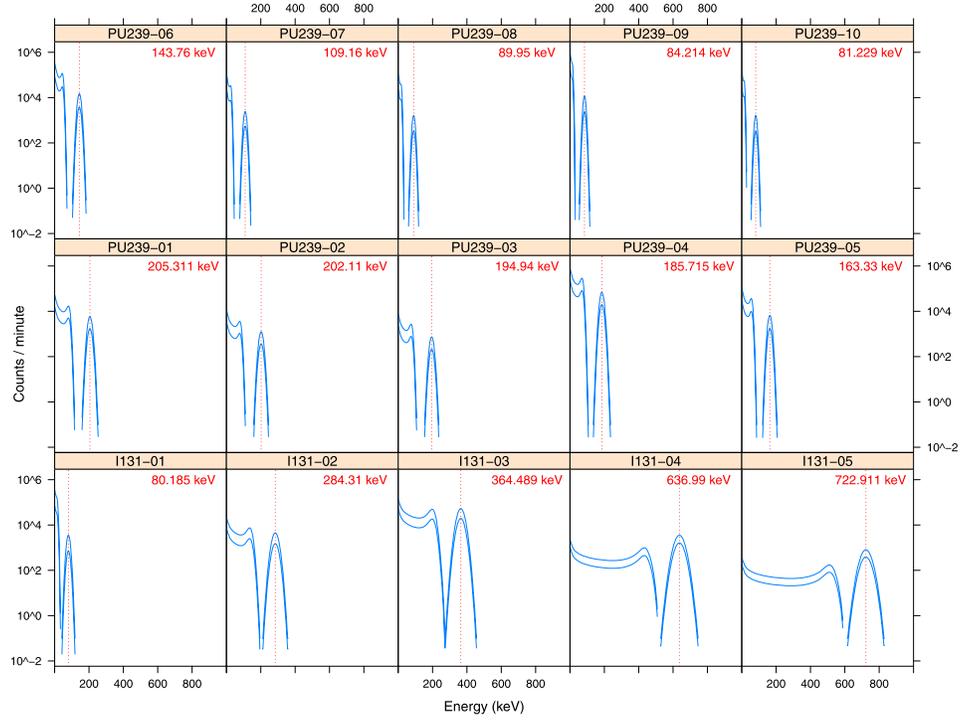}

\caption{Attenuated subspectra of monoenergetic gamma emissions for
$^{131}$I and $^{239}$Pu, on the logarithmic scale, whose branching
ratios are greater than $0.5\%$, under carbon shielding with $x = 0$
(i.e., no shielding, upper curves) and $x=10$ (lower curves).}
\label{fig:carbonSubspectra10Ch3}
\end{figure}

Figure~\ref{fig:leadSubspectra10Ch3} shows, on the logarithmic scale,
the attenuation of the mean subspectra of monoenergetic gamma emissions
for $^{131}$I and
$^{239}$Pu whose branching ratios are greater than $0.5\%$, contrasting
the case under lead shielding with $x = 10$ versus no shielding. In
particular, the upper curve in each sub-figure there is the detector
response function (on the logarithmic scale) to the monoenergetic gamma
emission from the nuclide. Figure~\ref{fig:carbonSubspectra10Ch3} shows
those under carbon shielding.
%

%

%
%
\begin{figure}

\includegraphics{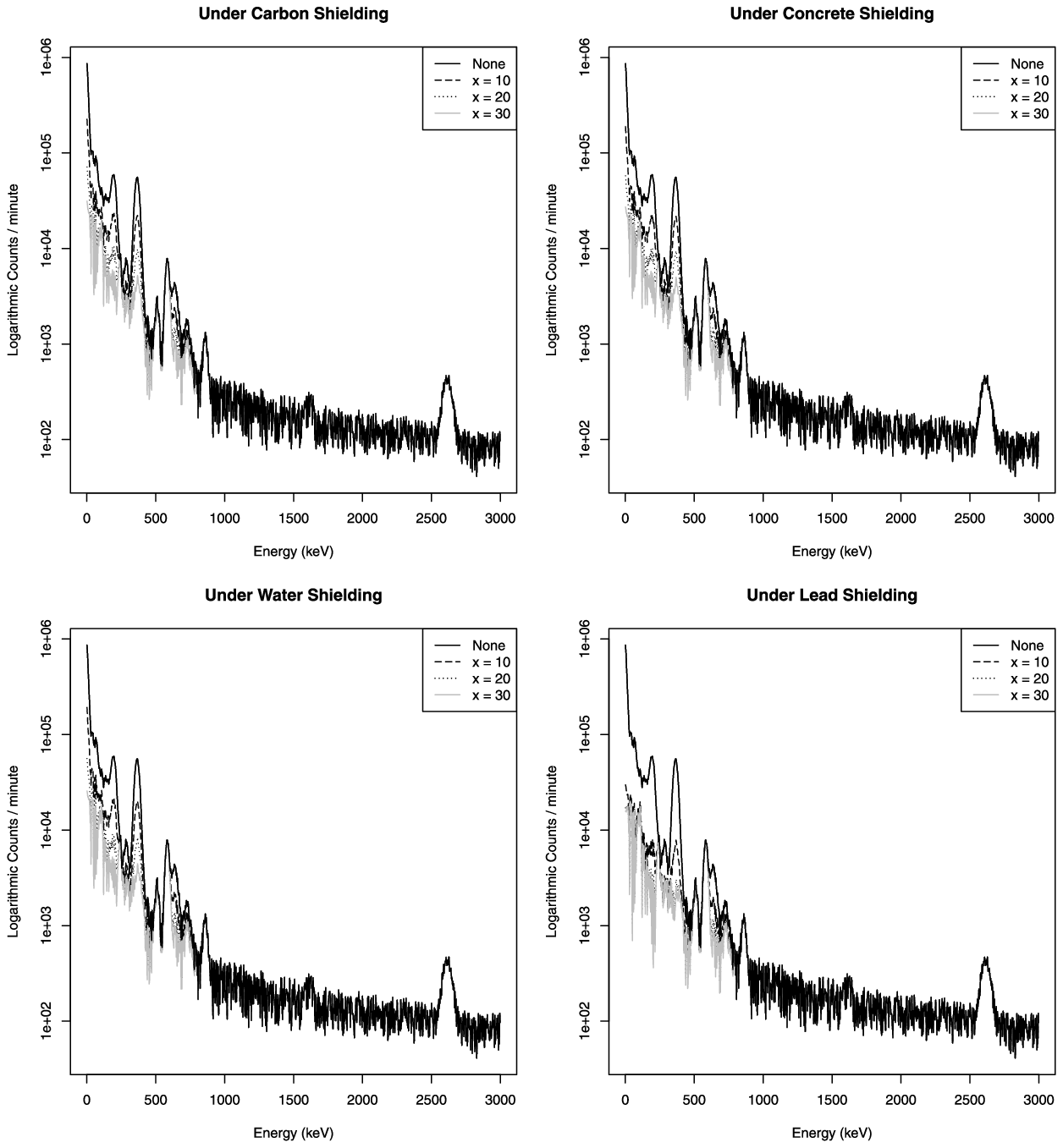}

\caption{Simulated signals under shielding where the mass thickness
$x$ is in g/cm$^2$.}
\label{fig:carbon10-20-30Ch3}
\end{figure}

%

A sample of simulated signals is presented in Figure~\ref{fig:carbon10-20-30Ch3} per intervening material.
Empirical power and size of both tests are based on 2000 replications,
with the empirical size checked against the nominal significance level
of 0.05. The LM test statistics were computed with the constrained ML
estimators obtained via the EM algorithm with the convergence criterion
of the relative maximum change in the $L^1$ norm of the parameter
estimate being less than $10^{-9}$, in all reported results; see the
supplemental article \citet{chan2013suppl}.

We first consider the case of carbon shielding.
In practice, the nature of the intervening material is unknown. So we
perform the LM test first assuming the intervening material is carbon
and then do the same test assuming one of three other misspecified
intervening materials, namely, concrete, lead and water. The derived
asymptotic $\chi^2$ distribution for the LM test allows us to
asymptotically calibrate the test, for example, by computing the approximate
$p$-value based on the $\chi^2$ distribution. Moreover, that the limiting
distribution is identical regardless of the true composition of
the sources (true $\mathbf{b}$ value) under the null hypothesis
justifies using the bootstrap for calibrating the LM test, which may be
more accurate for low signal-to-noise cases.

The empirical sizes of the LM test for various settings and under the
null hypothesis that $x = 0$ at a level of 0.05 are presented in Table~\ref{tab:I131Pu239Shield}. All sizes are close to the nominal size of
the test. An examination of quantile--quantile (q--q) plots (unreported)
of the LM test statistic against the theoretical Chi-square
distribution with one degree of freedom confirms the asymptotic
Chi-square distribution.
Note that the detection time is $\tau=1$ in all simulations, even
though the theoretical results are derived under
the assumption that $\tau\rightarrow\infty$. This suggests that the
results apply even for short times $\tau$.

\begin{figure}

\includegraphics{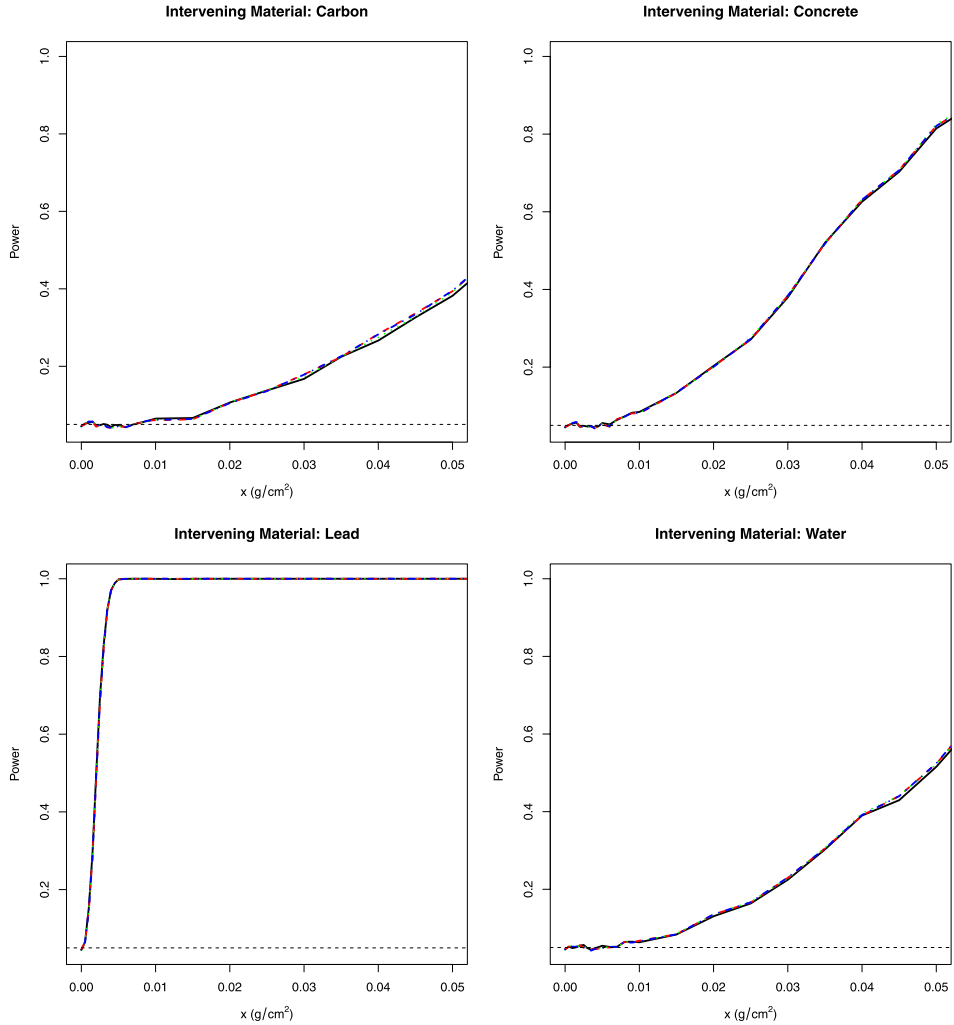}

\caption{Empirical power curves when the true intervening material is
carbon (top left), concrete (top right), water (bottom right) and lead
(bottom left) for the LM test with $x$ ranging from 0 to 0.05~g/cm$^2$.
The horizontal dashed lines show the 0.05 nominal level. The power
curves for testing for attenuation by carbon (concrete, lead, water)
are drawn as red dashed (green dotted, black solid, blue dot-dashed)
lines and nearly coincide in most cases.}
\label{fig:PowerCompare}
\end{figure}

The empirical power curves for the nominal 0.05 LM test for attenuation
due to carbon, water, concrete and lead are shown in Figure~\ref{fig:PowerCompare} when carbon is the true intervening material
(top left), when concrete is the true intervening material (top right),
when water is the true intervening material (bottom right), and when
lead is the true intervening material (bottom left plot). Each plot
gives power curves for $x$ ranging from 0 to 0.05 when the LM test is
administered at the nominal 0.05 significance level. All power curves
increase with increasing mass thickness, because greater mass thickness
entails stronger shielding, resulting in higher power. The power curves
for carbon, water and concrete shielding are of broadly similar shape.
However, the power curve for detecting lead shielding has a different
shape, as it approaches 100\% power rather quickly. The shape
difference in the power curves is due to the much stronger lead
attenuation function that has a somewhat different shape; see
Figure~\ref{fig:I131Pu239mac}.
We notice that the empirical power curves are almost identical for
different combinations of true and presumed intervening materials,
although the power is slightly decreased if the intervention material
is misspecified. In other words, the test is robust to the nature of
the intervening material, which likely owes to the fact that the
attenuation functions of these four intervening materials have rather
similar shapes.

\begin{table}
\caption{Empirical size of the LM test for shielding of Poisson signals
from $^{131}$I, 0.15 $^{239}$Pu and background, for each of 4
intervening materials}
\label{tab:I131Pu239Shield}
\begin{tabular*}{\textwidth}{@{\extracolsep{\fill}}lcccc@{}}
\hline
\textbf{Test} & \textbf{Carbon} &\textbf{Concrete} & \textbf{Lead} & \textbf{Water}\\
\hline
LM &0.046 &0.044 &0.046 & 0.045\\
\hline
\end{tabular*}
\end{table}

\subsection*{Composite intervening materials}

Next, we consider the case of two intervening materials. There are six
cases: carbon--lead shielding, concrete--lead shielding, water--lead
shielding, water--carbon shielding, water--concrete shielding and
concrete--carbon shielding.
The data simulation scheme is similar to the case of the simple
intervening material, with appropriate modifications as follows:
\begin{longlist}[1.]
\item[1.] Determine $c{}_{1lm}$ and $c{}_{2lm}$, where $c{}_{1lm}$ and
$c{}_{2lm}$ are the mass attenuation coefficients under the $m${th}, $m
= 1,2$ intervening material for each monoenergetic gamma emission of
$^{131}$I and $^{239}$Pu.
\item[2.] Find the mass thickness such that the empirical power is
approximately 50\% at the nominal 5\% LM test for each intervening
material alone. For each combination of two shielding materials, let
$x_M$ denote the vector of their mass thickness so determined. We vary
the vector of mass thickness as a multiple of $x_M$, with the multiple
being 0 to 1 with increment by $1/19$, in total 20 of them.
\item[3.] Calculate $a_{1l} = b_{1}  \tau\exp(-\sum_{m=1}^{2}c_{1lm}\times x_m)$
and $a_{2l} = b_{2}  \tau\times\break \exp
(-\sum_{m=1}^{2}c_{2lm}\times x_m)$ for each
$x_m, m=1,2$, respectively, and $\tau= 1$.
\item[4.] Simulate $Y_{i}$ as independent Poisson random variables with
mean\break $\sum_{j=1}^{J}\sum_{l=1}^{p_{j}}S_{ijl} a_{jl}$.
Note $J=3$ since we only have $^{131}$I, $^{239}$Pu and background.
\end{longlist}

We performed the LM test for shielding by the two true intervening
materials, but of unknown mass thickness. Then we repeated the LM test
assuming simple shielding by one of the two shielding materials. The
empirical sizes of the LM test for composite shielding by known
intervening materials at the nominal level of 0.05 are presented in
Table~\ref{tab:SizeComposite}. They are close to the nominal size of
the test, which is 0.05. In comparison, the empirical size is 0.046 for
testing for simple shielding by lead.
%
\begin{table}
\tabcolsep=0pt
\caption{Empirical sizes for the LM test when the Poisson signals from
$^{131}$I, 0.15 $^{239}$Pu and background are shielded by various
combinations of intervening materials}
\label{tab:SizeComposite}
\begin{tabular*}{\textwidth}{@{\extracolsep{\fill}}lcccccc@{}}
\hline\\[-2pt]
\textbf{Test} & \textbf{Carbon--Lead} &\textbf{Concrete--Lead} &
\textbf{Water--Lead}& \textbf{Water--Carbon}
&\multicolumn{1}{c}{\multirow{2}{40pt}[10pt]{\centering\textbf{Water--Concrete}}}&
\multicolumn{1}{c@{}}{\multirow{2}{40pt}[10pt]{\centering\textbf{Concrete--Carbon}}} \\
\hline
LM &0.050 &0.046 &0.048 & 0.051& 0.051&0.050\\
\hline
\end{tabular*}
%
\end{table}

\begin{figure}[b]

\includegraphics{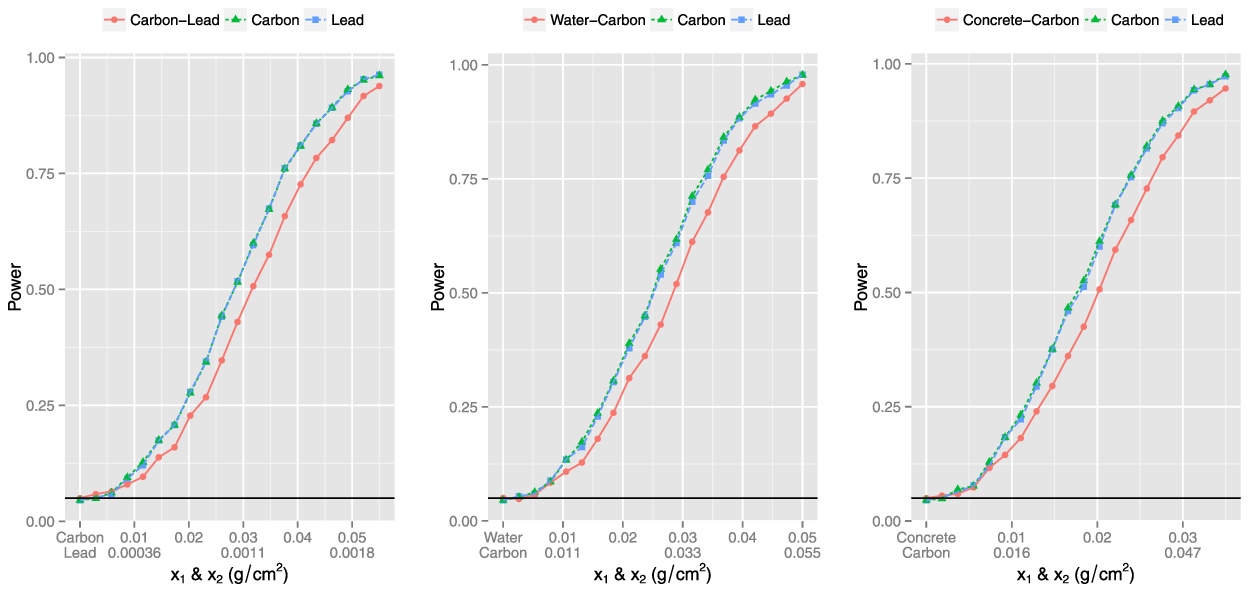}

\caption{Empirical power when the true intervening materials are carbon
and lead (left), water and carbon (middle) and concrete and carbon
(right). The power curves for testing for simple shielding by carbon
are drawn as green dashed lines, those of testing for simple shielding
by lead as blue longdash lines, and those for testing for composite
shielding by known materials as red solid lines.}
\label{fig:PowerCabonLead}
\end{figure}

\begin{figure}

\includegraphics{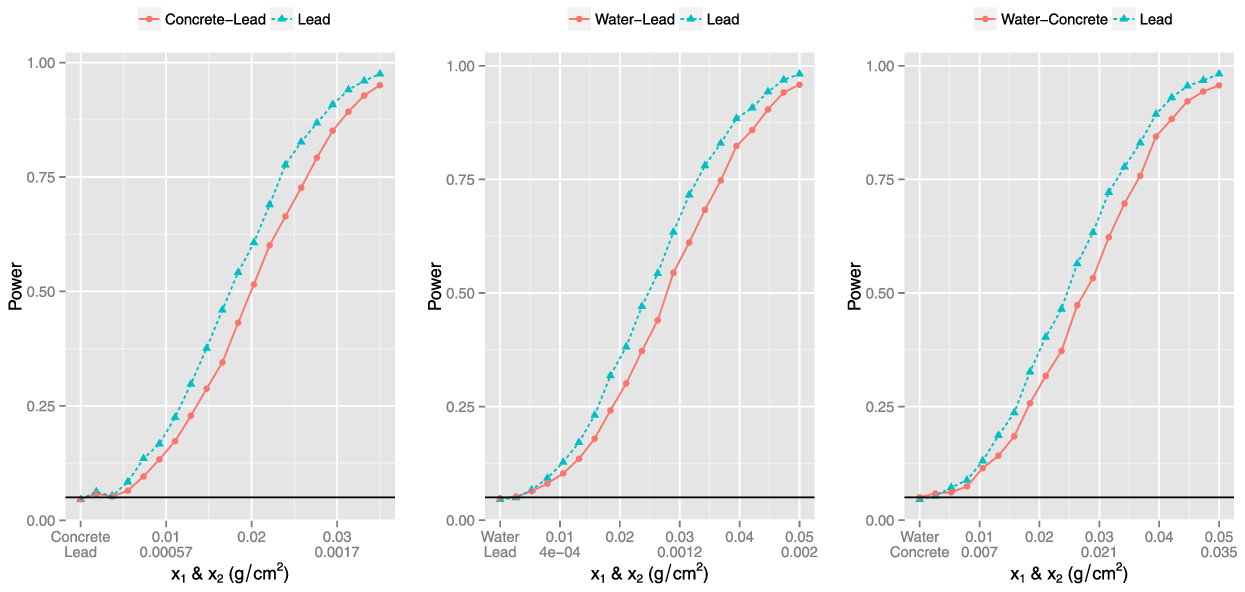}

\caption{Empirical power when the true intervening materials are
concrete and lead (left), water and lead (middle) and water and
concrete (right). The power curves for testing for simple shielding by
lead are drawn as blue dashed lines, and those for testing for
composite shielding by known materials as red solid lines.}
\label{fig:PowerConcreteLead}
\end{figure}

Figure~\ref{fig:PowerCabonLead} together with Figure~\ref{fig:PowerConcreteLead} contrast the empirical power of the test for
composite shielding by known intervening materials with those of tests
for simple shielding by possibly misspecified materials. These figures
suggest that, somewhat paradoxically, the LM test for simple shielding
may be more powerful than the LM test for composite shielding by known
intervening materials, even for tests for simple shielding by a
misspecified material.

\begin{figure}

\includegraphics{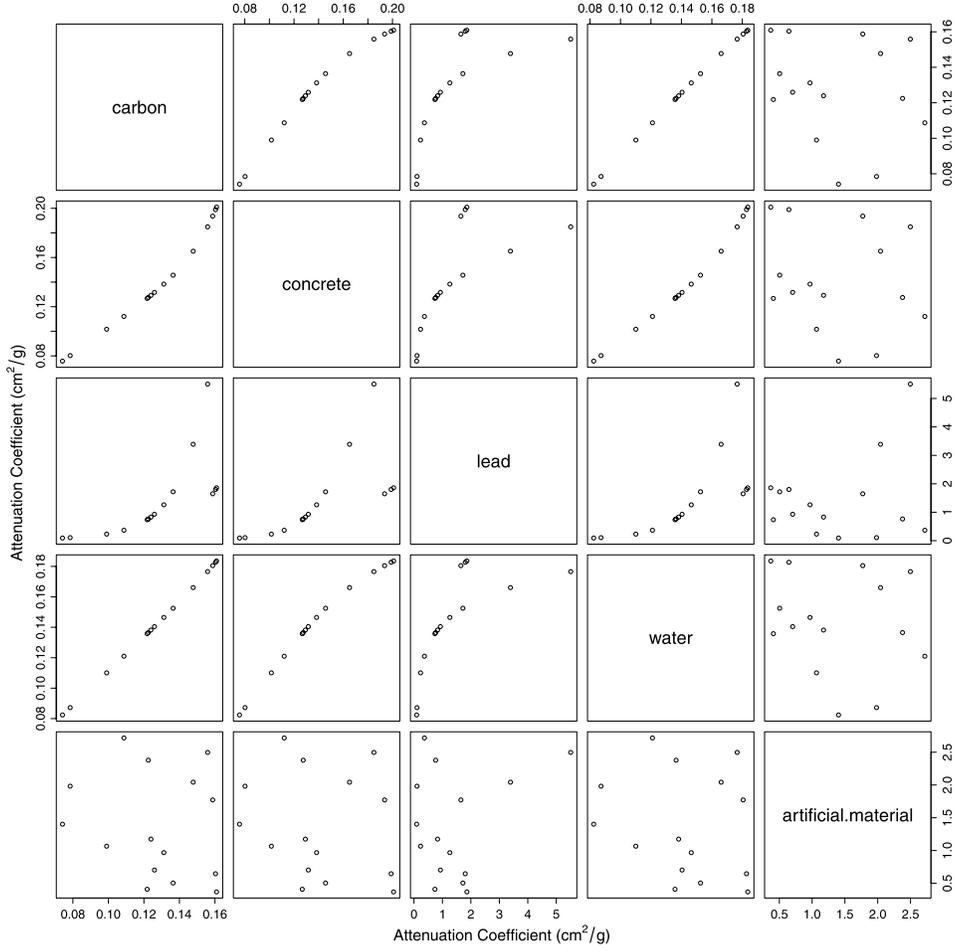}

\caption{Scatter plot of the mass attenuation functions of carbon,
concrete, water, lead and the artificial material.}
\label{fig:pairmacfake}
\end{figure}

\begin{figure}

\includegraphics{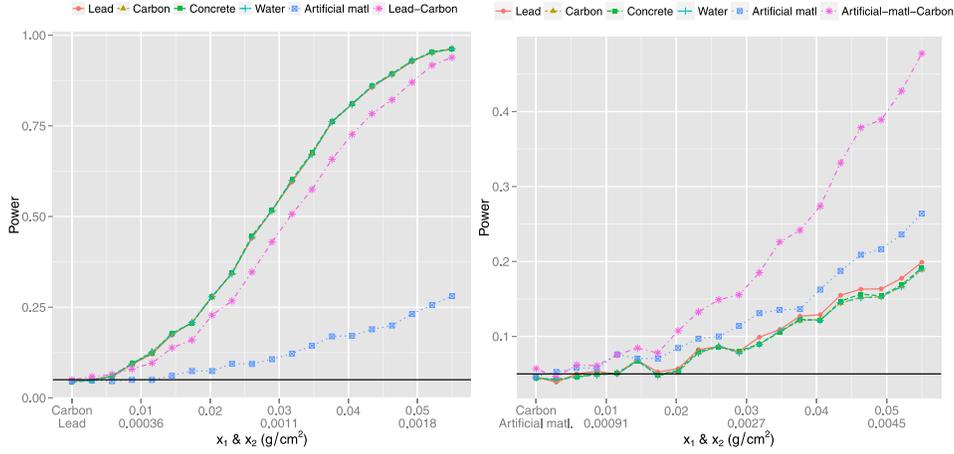}

\caption{Empirical power curves when the true intervening materials are
lead and carbon (left) and artificial material and carbon (right).
In both cases, the empirical power curves of testing for simple
shielding are almost the same, regardless of which intervening material
is assumed in the test.}
\label{fig:Powerfakelead}
\end{figure}

Below, we show that this phenomenon arises from the strong collinearity
between the attenuation functions of commonly used intervening materials.
Physically, the mass attenuation functions of intervening materials
are generally decreasing functions of the photon energy, so they tend
to be positively correlated.
Hence, a nonphysical mass attenuation function is used for illustrating
the problem of collinearity. For this purpose, we make up an
artificial, nonphysical material whose mass attenuation function is
``uncorrelated'' with those of the intervening materials considered so
far. We set the mass attenuation function of the artificial material to
be $\exp(\sin(\mathit{En}))$, where $\mathit{En}$ is the energy level of the photon. The
matrix of scatter plots between the mass attenuation functions of the
artificial material together with the other four intervening materials
is shown in Figure~\ref{fig:pairmacfake}, which displays strong
positive correlations among the attenuation functions of carbon,
concrete, water and lead. However, there appears to be no relationship
between the mass attenuation function of the artificial material and
those of the other four materials.
We then replicate the simulation experiment with signals simulated from
$^{131}$I, $0.15 ^{239}$Pu and background, that are shielded by the
artificial material and carbon, while we test for shielding by (i) the
artificial material and carbon, and (ii) by one of the 5 possible
intervening materials (carbon, lead, water, concrete and the artificial
material) alone. The mass thicknesses of the true composite intervening
materials are determined as before, with the maximum mass thickness of
the artificial material being 0.005. Figure~\ref{fig:Powerfakelead}
displays the empirical powers of these tests based on 2000
replications, which now reveals that the test for shielding with the
correct specification of the composite intervening materials
(artificial-material--carbon) has higher power than any test for
shielding by a single intervening material.

%

Multicollinearity can be regarded as a form of ill-conditioning in the
covariance matrix of the covariate. The condition number, which is
defined as the square root of the ratio of the largest eigenvalue to
the smallest eigenvalue of the covariance matrix, is often used to
diagnose multicollinearity. A matrix with a high condition number
(e.g., $>$30)
indicates the presence of ill-conditioning. The severity of the
multicollinearity increases with the condition number. The condition
number of the Fisher information matrix, $ {I_{11}(\tilde{{\bolds{\phi}
}})-I_{12}(\tilde{{\bolds{\phi} }})I^{-1}_{22}(\tilde{{\bolds{\phi}
}})I_{21}(\tilde{{\bolds{\phi} }})}$, for lead (mass thickness:
$9.47\times10^{-4}$) and carbon (mass thickness: $2.61\times10^{-2}$)
shielding is 185.8, while that of the artificial-material (mass
thickness: $2.37\times10^{-3}$) and carbon (mass thickness:
$2.61\times10^{-2}$) shielding is 12.88. This provides further
evidence that the counterintuitive phenomenon arises from strong
collinearity between the mass attenuation functions of commonly used
intervening materials.

\section{Sensitivity analysis}\label{sec:sensitivity}
In order to assess the impacts of errors in the detector response
functions (DRF) and background radiation on the proposed methods, we
repeat the simulation
study on the empirical power of the LM test for shielding of the source
($^{131}$I and $^{239}$Pu) by
carbon, with a similar setup as in the first simulation study in
Section~\ref{power}, but now with errors added to the DRFs and
background radiation. To mimic systematic errors, we generate data with
$S_{ijl}$ replaced by $\exp(m_{ijl}) \times S_{ijl}$, where, as
functions of $i$,
$\{ m_{ijl}, i=1,\ldots,N\}$ are jointly independent integrated random
walks. Specifically, $m_{i-2,j,l}-2 m_{i-1,j,l}+m_{i,j,l}=c\eta_i, i=N,
N-1,\ldots,1$, with the boundary
conditions that $m_{N+1,j,l}=m_{N+2,j,l}=0$, where $c>0$ is some scaling
factor to make a certain signal to noise ratio and the $\eta$'s are
independent standard normal variables. The integrated random walks
are initiated from the high end of the spectrum to mimic the situation
that systematic errors are largest in magnitude over the low end of the
spectrum. In particular, this approach can account for errors due to
neglecting Compton scattering by intervening materials, the modeling of
which is complex, as it
depends on the unknown geometric configuration encompassing the
source, the detector and intervening materials. We use two values of
$c$ that make the sample standard deviation of $\{m_{i j l},
i=1,2,\ldots, N\}$ equal to $0.00025$ and $0.00035$, respectively,
corresponding to roughly a maximum of 0.15\% and 0.21\% relative change
in the pulse height of the 16 DRFs for the two isotopes plus background
radiation, at each energy level.
Data were simulated with the randomly modulated DRFs and background,
but the tests assuming possible carbon shielding and correct source
radionuclides are carried out assuming no DRF or background errors.
Each experiment is replicated 6000 times.

\begin{figure}

\includegraphics{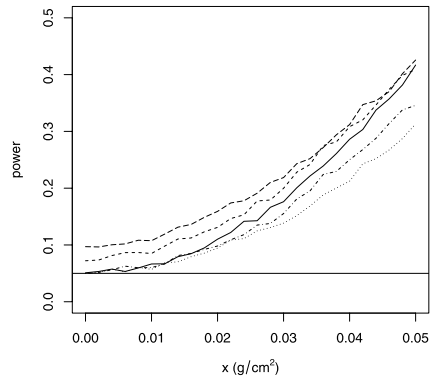}

\caption{Empirical power of the LM test for shielding $^{131}$I and $^{239}$Pu
by carbon with mass thickness $x$, with
data generated with the DRFs modulated by independent additive
integrated random walk errors on the logarithmic scale. Solid line:
$c=0$, that is, no DRF or background errors; dashed line: $c=0.00025$;
dotdash line: $c=0.00025$ with the test corrected to make the empirical
size equal to the 0.05 nominal size;
longdash line: $c=0.00035$; dotted line: $c=0.00035$ with the test
corrected to make the empirical size equal to the 0.05 nominal size.
Each experiment was replicated 6000 times.}\label{fig:error-in-DRF}
\end{figure}

Figure~\ref{fig:error-in-DRF} plots the empirical power curves for
$c=0, 0.00025$ and $0.00035$, which show that errors in the DRFs and
the background radiation inflated the size of the test, increasingly so
with the magnitude of $c$. The case of $c=0$ corresponds to no
shielding. The size-inflation
problem can be corrected by finding the cutoff that makes the size
of the test equal to the nominal 0.05. For example, for $c=0.00025$,
the 0.05 quantile of the test statistic under the null hypothesis of no
shielding, that is, $x=0$, is approximately $0.03234$, so the test can be
corrected by rejecting the null hypothesis if the $p$-value of the LM
test is less than 0.03504. Figure~\ref{fig:error-in-DRF} shows that the
size-correction procedure works at the
expense of reducing the power of the test, with larger errors
resulting in greater loss of power. Nonetheless, for small DFR and
background errors, the corrected test still has good power for
detecting shielding. An important research problem is to modify the
proposed methods to make them powerful and yet robust to errors in the
DRFs and background radiation.

%
\section{Discussion}\label{conclusion}
As noted, the LM test is attractive since only constrained maximum
likelihood estimation is needed. Simulation results suggest that the LM
test for simple shielding by lead is robust and powerful for testing
for simple or composite shielding by four common intervening materials,
even if the shielding materials do not contain lead. This result is
practically significant, as we generally do not know the number and
nature of the intervening materials. The usefulness of this robustness
result can be broadened by checking whether or not it continues to hold
with other shielding materials.

The problem of testing for shielding may also be addressed by a model
selection approach using some information criterion, with the selection
of the model with $x=0$ corresponding to no shielding. However, this
requires fitting a nonlinear model with positive $x$, which may result
in a
difficult global optimization problem, especially in the case of
composite shielding.

Implementing the proposed method in field applications, for example,
cargo screening, requires
further work to relax some restrictive assumptions of the proposed methods.
A main problem is that besides attenuating gamma rays, intervening
materials, especially those with low atomic number, such as carbon,
water and concrete, may cause down scattering of the gamma spectrum via
single or multiple external Compton scattering into the detector.
\citeauthor{Mitchell1989} [(\citeyear{Mitchell1989}), equations (11)--(13)]
proposed a semiparametric model
for modeling Compton scattering between a monoenergetic gamma ray and
materials outside the detector that enter into the spectrometer.
However, the full specification of the semiparametric model is specific
to the unknown geometry of the configuration encompassing the source,
the detector and the shielding materials. Alternatively, down
scattering may be studied by Monte Carlo techniques, for example, in
MCNP and GEANT, but this does not yield a closed-form solution for the
test statistics. Hence, down scattering due to Compton scattering
external to the spectrometer cannot be directly
incorporated in our proposed testing approach. While the impact of
Compton scattering into the detector may be of secondary importance,
its omission results in errors in the background radiation and may
reduce the power of the test, as demonstrated in Section~\ref{sec:sensitivity}.
Given the complexity needed for modeling external Compton scattering,
it seems that the problem may be best mitigated by modifying the
proposed methods to make them robust to errors in the background and
the DRFs.

While our main concern is to detect special nuclear material (SNM) and
so it is reasonable to consider only a few Pu isotopes and a few U
isotopes, it is imperative to reduce false alarms due to the presence
of other isotopes such as Neptunium (Np) and Americium (Am) that can be
weaponized, or naturally occurring radioactive materials (NORM) such as
granite, litter, etc. These ``nuisance'' isotopes and/or NORMs are less
likely shielded deliberately, the presence of each of which is, hence,
characterized by their unique overall DRF signature. The latter
signature DRFs can then be additively augmented into the Poisson
regression model, with their coefficients treated as nuisance
parameters in the tests. The theoretical properties and practical
performance of this approach for adjusting for nuisance radioactive
materials constitute an interesting future research direction.

The proposed methods assume known sources of SNMs which are, however,
generally unknown. One way to deal with this problem is to enumerate a
list of interesting combinations of source SNM, perform the test for
each such combination of source SNMs, and adjust for the multiplicity
of tests, for instance, by the Bonferroni rule or some other schemes
[\citet{Benjamini2010}].

Background radiation generally varies over time, so an interesting
problem is to devise a real-time updating scheme for background
radiation that incorporates concurrent covariate information, for
example, the vehicle type that may predict associated background
suppression by vehicles in cargo screening [\citet{lo2006}], and
to study the performance of the proposed methods using the background
updating procedure.

Another future research problem is to detect shielding in primary
screening with very low resolution gamma detectors.

\begin{supplement}[id=suppA]
\stitle{Proofs of Theorems \ref{Thm:chisq} and \ref{Thm:power} and
constrained maximum likelihood estimation via the EM algorithm}
\slink[doi]{10.1214/13-AOAS704SUPP} 
\sdatatype{.pdf}
\sfilename{aoas704\_supp.pdf}
\sdescription{This supplement contains (i) detailed proofs of Theorems
\ref{Thm:chisq} and \ref{Thm:power}, and (ii) an algorithm for constrained maximum likelihood
estimation assuming no shielding.}
\end{supplement}

%
%


\printaddresses

\end{document}